\documentclass[english]{revtex4}

\makeatletter

\usepackage{babel}

\makeatother
\begin{document}

\title{2+1 dimensional solution of Einstein Cartan equations}

\author{M. Horta\c csu, H.T. \"Oz\c celik, N. \"Ozdemir}


\affiliation{Istanbul Technical University, Faculty of Science and
Letters, 34469 Maslak, Istanbul, TURKEY}


\begin{abstract}
\noindent In this work a static solution of Einstein-Cartan (EC)
equations in 2+1 dimensional space-time is given by considering
classical spin-1/2 field as external source for torsion of the
space-time. Here, the torsion tensor is obtained from metricity
condition for the connection and the static spinor field is
determined as the solution of Dirac equation in 2+1 spacetime with
non-zero cosmological constant and torsion. The torsion itself is
considered as a non-dynamical field.
\end{abstract}

\maketitle

\section{Introduction}

According to the Einstein's theory of relativity, metric of
spacetime represents the gravitational field and the matter-energy
forms are related to spacetime curvature. In this theory,
spacetime has symmetric metric connection and it is called
pseudo-Riemannian spacetime due to its Lorentzian signature. Since the metricity condition
determines the connection uniquely in terms of metric components,
metric is the unique characteristic of Riemannian spacetime. Since
the Einstein field equations relate the geometry with considered
energy form, solution of the field equations will determine the
metric of spacetime for a given matter-energy form. The field
equations are non-linear and therefore, the solution metric may
not be the one, instead it may be a solution of a class of
solutions.

Existence of non-zero torsion of space-time is based on several
speculations in physics literature. For example, in references
\cite{hehl,hehl2}, it is pointed out that one of the reasons of
torsion is the spin of matter, which is coupled to the spacetime
via a non-Riemannian structure. It may exhibit a further influence
of geometry on  gravity.

On the other hand,  observations show that the present standard
model of particle physics  does not explain all of  the particle
spectrum. For example, neutrino oscillations can be result of a
more general theory \cite{maltoni}. To overcome this problem  new
fields or new symmetries can be introduced to theory at low
energies. Introducing torsion to the gravity theory may play a
role to explain these oscillations \cite{shapiro}. While the
metric is the only characteristic of the space-time in Einstein's
theory of gravity, in  theories with torsion the connection
coefficients are no longer symmetric and space-time is determined
by two independent characteristics, metric and torsion. Thus,
Einstein's relativity is generalized to the Einstein-Cartan
theory. Theories that include torsion are reviewed in details in
references \cite{shapiro,traut, pereira}\, where torsion is
considered as both a dynamical and a non-dynamical field.
Additional studies, specifying boundary conditions, include the
work of Peeters and Waldron \cite{PW}.

 In another line of
research, solutions to the Einstein's equations are investigated
in lower dimensions, since it is often easier to obtain solutions
on these manifolds. In 2+1 dimensions the gravitational wave does
not exist. The Weyl tensor vanishes identically and it is not
possible to have a Ricci flat space with non-trivial Riemann
tensor components. One has to introduce additional structure to
the model to get interesting solutions. Indeed Banados, Teitelboim
and Zanelli \cite{BTZ} introduced a nonvanishing cosmological
constant to obtain their black hole (BTZ) solution.  Deser, Jackiw
and Templeton \cite{DJT} studied the case with additional
topological terms (the DJT solution). Two of us have studied a
scalar field coupled to gravity in 2+1 dimensions in the past
\cite{HOY}. Torsion is added to the BTZ and DJT models in
references \cite{GHHM} and\cite{MP} respectively.  A recent work
of a model with torsion  is given by Blagojevi\' c and Cvetkovi\'
c \cite{BC} where a Maxwell field is coupled to gravity in 2+1
dimensions.

In this work, we consider torsion as a non-propagating field and solve the field
equations in the framework of the Einstein-Cartan theory. The source of torsion is an
external static radial classical spin-1/2 field. That is, the torsion is considered to be
coupled to the spinor field. An exact solution of the Einstein-Cartan field equations in
2+1 dimensions is obtained. Besides the torsion effect, the spinor field contributes to
the energy momentum tensor, right hand side of the Einstein equations, as a source and it
may also contribute to the spin (Cartan) equations. Furthermore, the Dirac equation
should be satisfied by spinor fields.  At the end of our calculations a solution of full
Einstein-Cartan and Dirac equations in the background of three dimensional space-time
with non-zero curvature and torsion, a metric for $U_3$ spacetime having the metricity
condition is obtained. The background spacetime is considered static and it is supposed
that the spacetime has positive cosmological constant.


We give mathematical definitions used throughout the work in the
following section and use the notation of \cite{yano}. Exact
solutions of Einstein-Cartan and Dirac equations are given in
section 3.

\section{Einstein-Cartan equations}
A measure of the space-time curvature is the difference between parallel transported
vectors along a closed path. In curved space-times, parallel transport of a vector along
a closed curve needs the introduction  of  the symmetric structure "the connection".
According to  Einstein's theory of relativity, the space-time has symmetric torsion free
Levi-Civita connection (or Riemannian connection) $\nabla$\,. If we allow the space-time
to have a metric tensor $g_{\nu\lambda}$\,, together with the symmetry condition , 
$\Gamma^\mu_{\nu\lambda}= \Gamma^\mu_{\lambda \nu}$, the metricity condition $\nabla_\mu\,
g_{\nu\lambda}=0$\, is
obtained. This symmetry condition determines the connection $\nabla_\mu$ uniquely. Then,
the connection coefficients, the Christoffel symbols, are given only in terms of metric
components
\begin{equation}
\Gamma^\mu_{\nu\lambda}={1\over 2}\,g^{\mu\sigma}\left({\partial_\nu\, g_{\nu\sigma}
}+{\partial_\lambda\, g_{\lambda\sigma} }-{\partial_\sigma\, g_{\nu\lambda}
}\right)\,,\label{chr3}
\end{equation}
where sub indices represent the partial derivatives with respect
to the coordinates. 

\noindent
When  torsion is taken into account, the
connection coefficients defined by (\ref{chr3})  no longer satisfy
the symmetry condition and space-time becomes non-Riemannian.
Thus, the Einstein's theory of gravity is generalized and modified
to the Einstein-Cartan (EC) theory. According to the EC theory,
affine connection is non-symmetric and the geometry can be
determined by two independent properties "the metric" and "the
torsion". 

\noindent
If
$\sim $  represents the quantities  written  in the
space-time with torsion, the torsion tensor $T$ is defined in
terms of the antisymmetric connection coefficients as follows:
\begin{equation}
T_{\nu\lambda}^{~~ \mu}=\tilde{\Gamma}_{\nu\lambda}^{ ~~ \mu}-{\tilde{\Gamma}_{
\lambda\nu}^{~~ \mu}}={ 2}\,\tilde{\Gamma}_{[\,\nu\lambda]}^{~~~\, \mu}\,.\label{torr}
\end{equation}
If we impose the metricity
condition,$\tilde{\nabla}_\mu\,g_{\nu\lambda}=0$, additionally, the
spacetime is called Riemann-Cartan and the metricity condition
enables us to define the connection in terms of metric and torsion
in a unique way
\begin{equation}
\tilde{\nabla}_\mu\,g_{\nu\lambda}=\partial_\mu\,g_{\nu\lambda}-{\tilde{\Gamma}_{\mu\nu}}^{~~
\rho}\,g_{\rho\lambda} -{\tilde{\Gamma}_{\mu\lambda}^{ ~~
\rho}}\,g_{\rho\nu}\,=0\label{connection}\,.
\end{equation}
In this equation $\tilde{\Gamma}_{\mu\nu}^{~~ \rho}$ represent
non-symmetric connection coefficients defined by
\begin{equation}
{\tilde{\Gamma}_{\nu\lambda}}^{~~ \mu}=\Gamma_{\nu\lambda}^{\mu}+K_{\nu\lambda}^{~~
\mu}\label{contor1}
\end{equation}
where ${\Gamma^\mu_{\nu\lambda}}$ are symmetric Riemannian
connection coefficients and $K$ is the contortion tensor
satisfying the relation $K_{ \nu\lambda}^{~~
\mu}=(T_{\nu\lambda}^{~~
\mu}+T^{\mu}_{\,\,\,\nu\,\lambda}+T^{\mu}_{\,\,\,\lambda\nu\,})/2$\,.
The raising and lowering indices are satisfied by the metric
tensor $g_{\mu\nu}$. From its definition it is easy to see that
the torsion tensor defined by (\ref{torr}) is anti-symmetric with
respect to first two indices and by using the metricity condition
given by (\ref{connection}) we get the result $K_{\mu\nu}^{~~
\rho}\,g_{\rho\lambda}+K_{\mu\lambda}^{~~ \rho}\,g_{\rho\nu}=0$\,,
or $K_{\mu\nu\lambda}+K_{\mu\lambda\nu}=0$ which means the
contorsion tensor is anti-symmetric with respect to its last two
indices.

\noindent The curvature tensor in spacetime with torsion is defined
in terms of spacetime connection coefficients
\begin{eqnarray}
&&{\tilde{R}}_{\nu\lambda\rho}^{\quad
\mu}=\partial_\rho\,\tilde{\Gamma}_{\lambda\nu}^{\,\,\,\,\,\, \mu}
-\partial_\lambda\,{\tilde{\Gamma}_{\rho\nu}}^{\,\,\,\,\,  \mu}
+{\tilde{\Gamma}_{ \rho\sigma}^{\,\,\,\,\,
\mu}}\,{\tilde{\Gamma}_{\lambda\nu}^{\,\,\,\,\,  \sigma}}
-{\tilde{\Gamma}_{ \lambda\sigma}^{\,\,\,\,\,
\mu}}\,{\tilde{\Gamma}_{\rho\nu}^{\,\,\,\,\, \sigma}}\,,
\end{eqnarray}
and the Ricci tensor and the Ricci scalar  are given by
\begin{eqnarray}
{\tilde{R}_{\,\mu\nu}}={{\tilde{R}}_{\sigma\nu\mu}^{~~~
\sigma}}\,,\qquad
\tilde{R}=g^{\mu\nu}\,{\tilde{R}_{\mu\nu}}\,.\label{curten}
\end{eqnarray}
Now, the field equations which have to be satisfied in space-times
with torsion  are called Einstein-Cartan equations
\begin{eqnarray}
&&{\tilde{R}_{\,\mu\nu}}-{1\over
2}({g_{\,\mu\nu}}\tilde{R}-2\Lambda)=\kappa\,t_{\mu\nu}\,,\label{energy}\\[.2cm]
&&T_{\nu\lambda}^{~~ \mu}-\delta^\mu_{\,\,\nu}T_{\rho\lambda}^{~~ \rho}
-\delta^\mu_{\,\,\lambda}T_{\nu\rho}^{~~ \rho}=\,\kappa\, S_{\nu\lambda}^{~~
\mu}\,.\label{entor}
\end{eqnarray}
Here $\kappa$ is the gravitational, $\Lambda$ is the cosmological constant and
$t_{\mu\nu}$  is the energy-momentum tensor. The equation (\ref{entor}) means that
torsion is related to the classical spin of the matter distribution $S_{\nu\lambda}^{~~
\mu}$. \noindent For brevity we will take $\kappa=1$ in our further calculations.

\section{2+1 dimensional solution with classical spin-1/2 source}
Consider the 2+1 dimensional space-time which has the following line element
\begin{equation}
ds^2=-v^2\,dt^2+w^2\,dr^2+r^2\,d\phi^2.\label{metric}
\end{equation}
Here $v\,,w$ are  radial coordinate $r$ dependent functions. Let
us consider a static external classical spin-1/2 field $\psi$
which is coupled to space-time as source of  torsion. Thus, the
considered space-time
has non-zero torsion.\\
If we impose the metricity condition (\ref{connection}) for the
metric (\ref{metric}), the non-zero connection coefficients
$\tilde{\Gamma}_{\mu\nu}^{~~ \rho}$ are obtained as follows:
\begin{eqnarray}
&&{\tilde{\Gamma}_{tr}}^{~\, t}={\tilde{\Gamma}_{rt}}^{~\, t}=\Gamma^t_{\,\,
rt}={v^\prime\over v}\,,\qquad {\tilde{\Gamma}_{tt}}^{~~ r}=\Gamma^{ r}_{\,\,
tt}={v\,v^\prime\over
w^2}\,,\nonumber\\[.2cm]
&&{\tilde{\Gamma}_{,rr}}^{~~ r}=\Gamma^r_{\,\,rr}={w^\prime\over w}\,,\qquad\qquad~~
{\tilde{\Gamma}_{r\phi}}^{~~ \phi}={\tilde{\Gamma}_{\phi r}}^{~~
\phi}=\Gamma^\phi_{\,\,r\phi}={1\over
r}\,,\nonumber\\[.2cm]
&& {\tilde{\Gamma}_{r\phi}}^{~~ t}=-{\tilde{\Gamma}_{\phi r}}^{~~ t}={u\,w^2\over
v^2}\,,\qquad~~{\tilde{\Gamma}^\phi_{\,\,rt}}=-{\tilde{\Gamma}_{tr}}^{~~ \phi}={uw^2\over
r^2}\,,\nonumber\\[.2cm]
&&{\tilde{\Gamma}_{\phi\phi}}^{~~ r}=-{r^2\over w^2}\,\tilde{\Gamma}_{r\phi}^{~~
\phi}\,,\qquad\qquad~~~ {\tilde{\Gamma}_{t\phi}}^{~~ r}=-{\tilde{\Gamma}_{\phi t}}^{~~
r}=u\,.
\end{eqnarray}
Here connection coefficients are obtained in terms of an arbitrary
function $u=u(r)$ which is to be determined by the field equations
and the Dirac equation. In these relations we see that $u(r)$
characterizes the torsion. As $u\rightarrow 0$ torsion becomes zero.
Then, the connection coefficients become symmetric and thus, the
Riemannian case is obtained.

\noindent Non-zero components of curvature tensor (\ref{curten}) are
\begin{eqnarray}
&&\tilde{R}_{tt}=-{1\over
r^2w^3}\,[2u^2w^5+rvv^\prime w^\prime-rvw(v^\prime+r^2vv^{\prime\prime})]\,,\nonumber\\[.2cm]
&&\tilde{R}_{t\phi}=-\tilde{R}_{\phi t}=u^\prime-u\left({1\over
r}+{v^\prime\over v}-{w^\prime\over w}\right)\,,\nonumber\\[.2cm]
&&\tilde{R}_{rr}={1\over
r^2 w^2w}\,[2u^2w^5+rvv^\prime w^\prime+r^2v^2w^\prime-r^2vv^{\prime}w^{\prime})]\,,\nonumber\\[.2cm]
&&\tilde{R}_{\phi\phi}={1\over v^2w^3}{2u^2w^5-rvwv^\prime+rv^2w^\prime}\,,
\end{eqnarray}
and the curvature scalar of the spacetime is
\begin{equation}
\tilde{R}={2\over
r^2v^2w^3}\left(3u^2w^5+rvw^\prime(v+rv^\prime)-rvw(v^\prime+rv^{\prime\prime})\right)\,,
\end{equation}
When $u$ is taken zero, the Riemannian situation is obtained. In the
presence of torsion, non-zero components of the totally
anti-symmetric torsion tensor, defined by (\ref{torr}), become
\begin{eqnarray}
&&T_{t\phi}^{~~ r}=2u\,,\qquad T_{r\phi}^{~~ t}={2u\,w^2\over v^2}\,,\qquad T_{rt}^{~~
\phi}={2uw^2\over r^2}\,.\label{torrr}
\end{eqnarray}
Since the torsion tensor is totally anti-symmetric, spin
distribution (\ref{entor}) belonging to matter field, takes the
form
\begin{eqnarray}
&&S_{t\phi}^{~~ r}=2u\,,\qquad S_{r\phi}^{~~ t}={2u\,w^2\over v^2}\,,\qquad S_{rt}^{~~
\phi}={2uw^2\over r^2}\,.\label{torrr}
\end{eqnarray}
In this work we consider classical spin-1/2 matter field, which is
source of the torsion. In addition to
contributions coming from ordinary matter forms, energy-momentum
tensor in (\ref{energy}), in general, also includes spin-torsion
interactions.  In 2+1 dimensional space-time, however, spin-torsion
interaction term becomes zero; therefore, it does not contribute to
$t_{\mu\nu}$ . Thus, energy momentum tensor $t_{\mu\nu}$ given in
(\ref{energy}) consists only of the spin-1/2 field matter part  and
it is defined in the torsion background by
\begin{eqnarray}
&&t^{matter}_{\mu\nu}={i\over
4}\left[\left(\bar{\psi}\gamma_\mu\tilde{\nabla}_\nu\psi+\bar{\psi}\gamma_\nu\tilde{\nabla}_\mu\psi\right)-
\left(\tilde{\nabla}_\nu\bar{\psi}\gamma_\mu\psi+\tilde{\nabla}_\mu\bar{\psi}\gamma_\nu\psi\right)\right]\,\label{tmatt}
\end{eqnarray}
where
\begin{eqnarray}
&&\tilde{\nabla}_\mu\psi=\partial_\mu\psi+\Gamma_\mu \psi\,,\qquad
\tilde{\nabla}_\mu\bar{\psi}=\partial_\mu\bar{\psi}-\Gamma_\mu \bar{\psi}\,.
\end{eqnarray}
Here $\Gamma_\mu$ is the spinor connection, $\bar{\psi}$ is conjugate of the spinor field
$\psi$ having the relation $\bar{\psi}=\psi^\dagger\gamma^0$\,.

In an orthonormal frame the metric tensor of space-time can be
expressed in the following way:
\begin{eqnarray}
g_{\mu\nu}=e^a_{\,\mu}\,e^b_{\,\nu}\,\eta_{ab}\,,\quad
e^a_{\,\mu}\,e^\nu_{\,a}=\delta^\nu_\mu
\end{eqnarray}
where   $\eta_{ab}=(-,+,+)$ is 2+1 Minkowski metric.  Raising and
lowering of indices are made by using the metric tensor
$g_{\mu\nu}$,
\begin{equation}
e_{\mu a}=g_{\mu\rho}\,e^\rho_{\,a}\,.
\end{equation}
The spinor connections $\Gamma_\mu$ are defined as follows
\begin{equation}
\Gamma_\mu={1\over
8}\,[\gamma^a,\gamma^b]\,e^\nu_{\,a}\left(\partial_\mu\,e_{\nu
b}-{\tilde{\Gamma}_{\mu\nu}}^{~~\rho}\,e_{\rho b}\right)\,,
\end{equation}
where $\gamma^a$'s ($a=0,1,2$)\,, are 2+1 dimensional Minkowski gamma matrices
\begin{eqnarray} \gamma^0=\left(\begin{array}{cc}
0&1   \\
 1&0 \\
\end{array}\right)\,,\quad
\gamma^1=\left(\begin{array}{cc}
i&0   \\
 0&-i \\
\end{array}\right)\,,\quad
\gamma^2=\left(\begin{array}{cc}
0&-1   \\
 1&0 \\
\end{array}\right)\,
\end{eqnarray}
such that $\{\gamma^a,\gamma^b\}=-2\eta^{ab}$, and curved space-time gamma matrices
$\gamma^\mu$ can be obtained from flat space gamma matrices by means of orthonormal base
transformation $\gamma^\mu=e^\mu_{\,a}\,\gamma^a$\, satisfying the relation
$\{\gamma^\mu,\gamma^\nu\}=-2g^{ab}$. Spinor fields in 2+1 dimensions are given in terms
of their components as
\begin{eqnarray}
\psi=\left(\begin{array}{c}
\psi_1   \\
  \psi_2 \\
\end{array}\right)\,,\qquad
\bar{\psi}=\left( \bar{\psi}_1~~ \bar{\psi}_2 \right)=\left( {\psi}_2^*~~ {\psi}_1^*
\right)\,
\end{eqnarray}
where  "*" represents the complex conjugate of the functions. In
this problem spinor field $\psi$ and conjugate spinor field
$\bar\psi$ are considered static $r$ dependent functions only. An
orthonormal base can be chosen in the following form for the
spacetime (\ref{metric})
\begin{equation}
e^\mu_{\,a}=\left(\begin{array}{ccc}
1/v&0&0   \\
0&1/w&0 \\
0&0&1/r\\
\end{array}\right)\,,\quad
e^a_{\,\mu}=\left(\begin{array}{ccc}
v&0&0   \\
0&w&0 \\
0&0&r\\
\end{array}\right)\,.
\end{equation}
Then, 2+1 dimensional curved space-time gamma matrices
$\gamma^\mu$\, become
\begin{eqnarray} \gamma^{\,t}=\left(\begin{array}{cc}
0&1/v   \\
1/v&0  \\
\end{array}\right)\,,\quad
\gamma^{\,r}=\left(\begin{array}{cc}
i/w&0   \\
0 &-i/w \\
\end{array}\right)\,,\quad
\gamma^\phi=\left(\begin{array}{cc}
0&  -1/r \\
1/r &0 \\
\end{array}\right)\,.
\end{eqnarray}
Non-zero components of the Einstein tensor defined by (\ref{energy}) for the metric
(\ref{metric}) and torsion tensor (\ref{torrr}) will satisfy the field equations
\begin{eqnarray}
&&G_{tt}-t_{tt}=-\Lambda\,v^2+{u^2w^2\over r^2}+{v^2w^\prime\over w^2}\,,\\[.2cm]
&&G_{tr}-t_{tr}=G_{rt}-t_{rt}=-{i\over4}(-v(\bar{\psi}_2\psi_1^\prime+\bar{\psi_1}\psi_2^\prime)+\psi_2(v^\prime\bar{\psi}_2+v\bar{\psi}_1^\prime)+
\psi_1(v^\prime\bar{\psi}_1+v\bar{\psi}_2^\prime)\,,
\\[.2cm]
&&G_{t\phi}-t_{t\phi}=-G_{\phi t}+t_{\phi t}=u^\prime-u\left({1\over r}-{v^\prime\over v}-{w^\prime\over w}\right)\,,\label{fieldeq1}\\[.2cm]
&&G_{rr}-t_{rr}=\Lambda w^2-{u^2w^4\over r^2v^2}+{v^\prime\over r v}+{1\over
2r^2v^2}\left(-rvuw^3(\psi_1\bar{\psi}_1+\psi_2\bar{\psi}_2)+2rvv^\prime+r^2v^2w(\psi_1^\prime\bar{\psi}_1-\psi_2^\prime\bar{\psi}_2-\psi_1\bar{\psi}_1^\prime+\psi_2\bar{\psi}_2^\prime)\right),
\\[.2cm]
&&G_{r\phi}-t_{r\phi}=G_{\phi r}-t_{r\phi }={i\over
4v}\left(uw^2(\psi_1\bar{\psi}_1+\psi_2\bar{\psi}_2)+rv(-{\psi}_1^\prime\bar{\psi_2}+{\psi}_2^\prime\bar{\psi_1}-\bar{\psi}_1^\prime\psi_2+\psi_1\bar{\psi}_2^\prime)\right)\,,
\\[.2cm]
&&G_{\phi\phi}-t_{\phi\phi}=\Lambda r^2-{u^2w^2\over v^2}+{r^2\over
vw^3}\,(w^{\prime\prime}-v^\prime w^\prime)\,.\label{fieldeq}
\end{eqnarray}
Here $^\prime$ denotes the derivative with respect to $r$.

A set of solutions to 2+1 dimensional field equations can be found by solving
differential equations (\ref{fieldeq1}, \ref{fieldeq}), but solution of the Dirac
equation to be satisfied by spinor field will restrict us to a specific solution.

In general, the 2 dimensional spinor fields satisfy the following Dirac equations:
\begin{eqnarray}
&&i\gamma^\mu\tilde{\nabla}_\mu\psi-m\psi=0\,,\qquad
i\tilde{\nabla}_\mu\bar{\psi}\gamma^\mu+m\bar{\psi}=0
\end{eqnarray}
with mass $m$\, of the spinor field.

Dirac equations in this background become
\begin{eqnarray}
&&i\gamma^\mu\tilde{\nabla}_\mu\psi-m\psi=i\gamma^\mu(\partial_\mu\psi+\Gamma_\mu
\psi)-m\psi=\left(\begin{array}{c} \displaystyle{1\over 2 r v
w}{\left[uw^2(\psi_1-\psi_2)-rv^\prime\psi_2-2rv(m w\psi_1+\psi_1^\prime)\right]}\\[.2cm]
\displaystyle{1\over 2 r v w}{\left[uw^2(-\psi_1+\psi_2)+rv^\prime\psi_1-2rv(mw\psi_2-\psi_2^\prime)\right]}\\
\end{array}\right)=0\,,\\[.2cm]
&&i\tilde{\nabla}_\mu\bar{\psi}\gamma^\mu+m\bar{\psi}=(\partial_\mu\bar{\psi}-\Gamma_\mu
\bar{\psi})\gamma^\mu+m\bar{\psi} \nonumber\\[.2cm]
&&=\left(\displaystyle{1\over
2rvw}{\left[-uw^2(\bar{\psi_1}+\bar{\psi_2})+rv^\prime\bar{\psi}_2+2rv(mw\bar{\psi}_1-\bar{\psi}_1^\prime
)\right]}\quad \displaystyle{1\over
2rvw}{\left[-uw^2(\bar{\psi_1}+\bar{\psi_2})-rv^\prime\bar{\psi}_2+2rv(mw\bar{\psi}_1+\bar{\psi}_1^\prime
)\right]}\right)=0\,\nonumber\\[.2cm]
\end{eqnarray}
for the  metric given in (\ref{metric})\,. And a solution of Einstein-Cartan equations
can be found as
\begin{eqnarray}
&&v=r\,,\qquad  w={1\over r\sqrt{c^2-\Lambda}}\,,\qquad u=cr^3\sqrt{c^2-\Lambda}\,,
\end{eqnarray}
\begin{eqnarray}
&&\psi_1=\bar{\psi_2}=d_1\cos{\sqrt{\Lambda-c^2+4m(c-m)}\over 2\sqrt{c^2-\Lambda}}\log r+d_2\sin{\sqrt{\Lambda-c^2+4m(c-m)}\over 2\sqrt{c^2-\Lambda}}\log r\,\nonumber\\[.7cm]
&&\psi_2={1\over
c+\sqrt{c^2-\Lambda}}\left[d_3(c-2m)-d_4(\sqrt{\Lambda-c^2+4m(c-m)})\right]\cos{\sqrt{\Lambda-c^2+4m(c-m)}\over
2\sqrt{c^2-\Lambda}}\log r\nonumber\\[.2cm]
&&\qquad+{1\over
c+\sqrt{c^2-\Lambda}}\left[d_4(c-2m)-d_3(\sqrt{\Lambda-c^2+4m(c-m)})\right]\sin{\sqrt{\Lambda-c^2+4m(c-m)}\over
2\sqrt{c^2-\Lambda}}\log r
\,,\nonumber\\[.7cm]
&&\bar{\psi}_2={1\over
c+\sqrt{c^2-\Lambda}}\left[-d_3(c-2m)+d_4(\sqrt{\Lambda-c^2+4m(c-m)})\right]\cos{\sqrt{\Lambda-c^2+4m(c-m)}\over
2\sqrt{c^2-\Lambda}}\log r\nonumber\\[.2cm]
&&\qquad+{1\over
c+\sqrt{c^2-\Lambda}}\left[-d_4(c-2m)+d_3(\sqrt{\Lambda-c^2+4m(c-m)})\right]\sin{\sqrt{\Lambda-c^2+4m(c-m)}\over
2\sqrt{c^2-\Lambda}}\log r \,\\[.2cm]\nonumber
\end{eqnarray}
with $d_1, d_2$ real,$d_3, d_4$ imaginary constants. Furthermore,  $c>\sqrt{\Lambda}$ is
a positive constant and subject to the constraint $4m(c-m)>c^2-\Lambda$. Then, the metric
(\ref{metric}) takes the form
\begin{equation}
ds^2=-r^2\,dt^2+{1\over (c^2-\Lambda)r^2}\,dr^2+r^2d\phi^2\,.\label{solmet}
\end{equation}
In this solution the non-zero curvature tensor components are
\begin{eqnarray}
&&\tilde{R}_{tt}=-2r^2\Lambda\,,\qquad \tilde{R}_{rr}={2\Lambda\over
r^2(c^2-\Lambda)}\,,\qquad \tilde{R}_{\phi\phi}={2 r^2\Lambda}
\end{eqnarray}
and the scalar curvature for (\ref{solmet}) becomes
$\tilde{R}=6\Lambda\,$.  We note that it is essential to have a
non-vanishing cosmological constant to have a metric with non-zero
curvature.

It is also worthwhile to mention that the Laplace operator does not change its form for
the torsion spacetime with metric (\ref{solmet}). The Laplace operator in torsion
spacetime is
\begin{equation}
\tilde{\nabla}_\mu\tilde{\nabla}^\mu={\nabla}_\mu{\nabla}^\mu-K_{\mu\nu}^{\,\,\,\,\,\mu}g^{\nu\sigma}\partial_\sigma=
{1\over\sqrt{-g}}\partial_\mu(\sqrt{-g}g^{\mu\nu}\partial_\nu)-K_{\mu\nu}^{\,\,\,\,\,\mu}g^{\nu\sigma}\partial_\sigma
\end{equation}
where $K_{\mu\nu}$ is the contorsion tensor defined by (\ref{contor1}). Since the
contorsion tensor is zero for the spacetime  (\ref{solmet}) having torsion (\ref{torrr}),
the Laplace equation reduces to Riemannian case.

\section{Conclusion}

By considering Dirac type static classical spin-1/2 field as the source of torsion of the
space-time, a solution of 2+1 dimensional Einstein-Cartan equations is obtained. Here,
torsion is taken a non-propagating field and couples to the spacetime via spinor field.
In the beginning, we started with a general type of the spinor, it could either be Dirac
or Majorana. But, during the calculations we see that the torsion tensor is determinative
of the type of the spinor and only Dirac type source fulfills the equations.Therefore,
the Dirac equation determines the final form of the metric.

In the further works, it will be interesting to work with 3+1 and higher dimensional
spacetime and  to examine  whether the
torsion tensor forms a certain structure and restricts the
type of the spinor field to the Dirac or Majorana type in general. 

\vspace{.7truecm}

{\large\noindent \bf{Acknowledgements}} We dedicate this work to
Prof. Dr. Yavuz Nutku on his 65th birthday. This work is partially
supported by TUBITAK, the Scientific and Technological Council of
Turkey. The work of M.H. is also supported by TUBA, Academy of
Sciences of Turkey.

\noindent

\begin{thebibliography}{9}

\bibitem{hehl} W.F. Hehl, P. von der Heyde, G.D. Kerlick,  Phys. Rev. D{\bf 10}(1974) 1066-1069,
W.F. Hehl, P. von der Heyde, G.D. Kerlick,  Rev. of Mod. Phys. {\bf 48}(1976) 393-416

\bibitem{hehl2} W.F. Hehl, P. von der Heyde,  Phys. Rev. D{\bf 10}(1973) 179-196,


\bibitem{maltoni} M.C. Gonzalez-Garcia, M. Maltoni,
Phys. Rept. {\bf 460}(2008) 1-129, arXiv:0704.1800 (hep-ph),

\bibitem{shapiro} I.L. Shapiro, Phys. Rept. 357(2002) 113, hep-th/0103093

\bibitem{traut} A. Trautman,  2006, {\it Encyclopedia of Mathematical Phys.} ed.: J.P. Fran\c coise, G.L. Naber, S.T.
Tsou, Oxford: Elsevier, vol.2, 189-195, gr-qc/0606062

\bibitem{pereira} H.I. Arcos, J.G. Pereira, Int. J. Mod. Phys.D{13}(2004)2193-2240

\bibitem{PW} K.Peeters and A. Waldron, JHEP 9902, (1999) 24

\bibitem{BTZ} M. Banados, C. Teitelboim and J. Zanelli, Phys. Rev.
Lett. {\bf 69} (1992) 1849-51

\bibitem{DJT} S. Deser, R. Jackiw and S. Templeton, Ann. Phys.
(N.Y.)  {\bf 140} (1982)372-411

\bibitem{HOY} M. Horta\c csu, H.T. \" Oz\c celik and B. Yap\i\c
skan, Gen. Rel. and Grav. {\bf 35} (2003) 1209-1221

\bibitem{GHHM} A.Garcia, F.W.Hehl, C. Heinecke and A. Macias,
Phys. Rev. {\bf 67} (2003) 124016

\bibitem{MP} E.W. Mielke and P. Baekler, Phys. Lett. {\bf A 156}
(1991) 399-403

\bibitem{BC}  M. Blagojevi\' c and B. Cvetkovi\' c, gr-qc
0804.1899

\bibitem{yano} K. Yano, 1965, {\it Differential Geometry on Complex and
Almost Complex Spaces}, Pergamon Press



\end{thebibliography}
\end{document}